# Spatially resolving unconventional interface Landau quantization in a graphene monolayer-bilayer planar junction

Wei Yan[§], Si-Yu Li[§], Long-Jing Yin, Jia-Bin Qiao, Jia-Cai Nie, and Lin He*

**Graphene hybrid planar structures consisting of two regions with different quantum Hall (QH) states exhibit unusual transport properties[1-5], originating from chiral edge states equilibration at the interface of the two different regions[6]. Here we present a sub-nanometre-resolved scanning tunnelling microscopy (STM) and spectroscopy (STS) study of a monolayer-bilayer graphene planar junction in the QH regime. The atomically well-defined interface of such a junction allows us to spatially resolve the interface electronic properties. Around the interface, we detect Landau quantization of massless Dirac fermions, as expected in graphene monolayer, below the charge neutrality point $N_c$ of the junction, whereas unexpectedly, only Landau quantization of massive Dirac fermions, as expected in graphene bilayer, is observed above the $N_c$. The observed unconventional interface Landau quantization arises from the fact that the quantum conductance across the interface is solely determined by the minimum filling factors (number of edge modes) in the graphene monolayer and bilayer regions of the junction[6,7].**

Department of Physics, Beijing Normal University, Beijing, 100875, People's Republic of China
[§]These authors contributed equally to this work.
* Email: helin@bnu.edu.cn

There are two general methods to realize graphene hybrid QH structures. The first method is based on locally gated graphene monolayer or bilayer, where local gates are used to individually control carrier densities and/or carrier types in two adjacent regions[1-6,8,9], thus creating different QH states with different filling factors in a magnetic field. The second one is to use graphene monolayer and bilayer hybrid planar structures[7,10-12], where two adjacent regions with different QH states exist naturally because of the different Landau level sequences in the monolayer[13-16] and bilayer[17-19] regions. The atomically well-defined interface of the graphene monolayer-bilayer planar junction, which is fixed at the edge of the bilayer lattice, provides unprecedented opportunity to spatially explore the electronic properties of the interface in the QH regime.

Figure 1a shows an atomic-resolution topographic STM image of a graphene monolayer-bilayer planar structure (see Supplementary Information A.1 for details of STM measurement). The graphene bilayer was grown on Rh foil via a traditional ambient pressure chemical vapour deposition method[20,21] (see Supplementary Information A.2 for details). To realize the monolayer-bilayer hybrid structure, we used the thermal strain engineering based on mismatch of thermal expansion coefficients between the graphene and the metallic foil[22,23]. For graphene monolayer grown on metallic foils, the thermal strain engineering could lead to the formation of nanoscale periodic graphene ripples, which are beyond the descriptions of continuum mechanics[22,23]. For the graphene bilayer studied here, similar thermal annealing of the sample (see Supplementary Information A.3 for details) not only generates nanoscale quasi-periodic ripples (Fig. 1a and also see Supplementary Fig. S1), but also results in a completely decoupling of the adjacent two layers in the rippled region. The periodic protuberances along the ripples may form as a result of a twist between the adjacent two layers[20,21,24], which is generated by their slight difference of the out-of-plane rippling. The atomic-resolution STM images of the rippled region exhibit clear honeycomb lattice, which can be regarded as a signature that the top layer is a graphene monolayer decoupled from the bottom layer. Only a small part of the sample still retains the stacking order of Bernal bilayer, as revealed by the triangular lattice in the relatively flat region (Fig. 1a). According to the previous study[22], the graphene is suspended above, that is, not in direct contact to the Rh substrate in the rippled areas, therefore, it is allowed to relax the compressive strain through rippling. In the relatively flat region, the graphene bilayer is directly pinned to the Rh foil, preventing the out-of-plane rippling. According to our STM measurement (the inset of Fig. 1a), an interface between the graphene bilayer (triangular lattice) and the monolayer (honeycomb lattice) regions can be sharply defined down to the atomic scale.

The local Landau quantization of such a monolayer-bilayer junction was measured by STS at 4.5 K using standard lock-in techniques. STS spectra recorded in the rippled region (~ 10 nm away from the interface) with different magnetic fields show Landau quantization of massless Dirac fermions, as expected to be observed in a graphene monolayer (Fig. 1b). The observed Landau Level (LL) energies $E_n$ depend on the square-root of both level index $n$ and magnetic field $B$[13-16]

$$E_n = \text{sgn}(n)\sqrt{2e\hbar v_F^2 |n| B} + E_0, \qquad n = ...-2,\ -1,\ 0,\ 1,\ 2... \qquad (1)$$

Here $E_0$ is the energy of Dirac point, $e$ is the electron charge, $\hbar$ is the Planck's constant, and $v_F$ is the Fermi velocity. The linear fit of the experimental data to Eq. (1), as shown in Fig. 1d, yields a Fermi velocity of $v_F = (1.208 \pm 0.016) \times 10^6$ m/s for electrons and $v_F = (0.992 \pm 0.014) \times 10^6$ m/s for holes. Similar large electron-hole asymmetry, which was attributed to the enhanced next-nearest-neighbor hopping enabled by the twist, has been observed previously in twisted graphene bilayers[24]. The observation of LL sequence of the massless Dirac fermions, as shown in Fig. 1b and 1d, demonstrates the efficient decoupling of the surface graphene monolayer from the bottom layer in the rippled region.

In the bilayer region, the tunnelling spectra exhibit Landau quantization of massive Dirac fermions (Fig. 1c). The substrate (here the Rh foil) could break the symmetry of the two adjacent layers and generate a finite gap in the parabolic bands of the Bernal bilayer[19,24-26]. With increasing magnetic field, two valley-polarized quartets, $LL_{(0,+)(1,+)}$ and $LL_{(0,-)(1,-)}$, are generated at the conduction band edge and valence band edge, respectively (here 0 and 1 are the LL indices and the symbols +/- are valley indices). The quartet $LL_{(0,+)(1,+)}$ is mainly localized on the topmost graphene layer and the other quartet $LL_{(0,-)(1,-)}$ resides mainly on the second layer (see Supplementary Information A.4 for discussion about the different charge neutrality points in the monolayer and bilayer regions). Therefore, the signal of the quartet $LL_{(0,+)(1,+)}$ is much stronger in the spectra (Fig. 1c). The energy gap $E_g$ of the bilayer is estimated to be about 80 meV, which is consistent well with the range of values reported previously for Bernal bilayer on various substrates[19,25-29]. Besides the two field-independent valley-polarized quartets, the other LLs show linear field dependence[17-19]:

$$E_n = \pm\left[\hbar\omega_c(n(n-1))^{1/2} + E_g/2\right], \qquad n = 0,1,2... \qquad (2)$$

Here $\omega_c = eB/m^*$ is the cyclotron frequency, $m^*$ is the effective mass of quasiparticles. The effective mass for both electrons and holes are estimated to be $(0.013 \pm 0.001)m_e$ (here $m_e$ is the free-electron mass). The relatively smaller effective mass of the quasiparticles comparing with that reported previously in the graphene Bernal bilayer[17-19] is attributed to a much weaker interlayer hopping in our sample (see Supplementary Fig. S2 and A.5 for discussion). Therefore, the STS spectra together with the STM measurements, as shown in Fig. 1, provide direct and compelling evidence that the structure in Fig. 1a is a graphene monolayer-bilayer planar junction.

Figure 2a shows tunnelling spectra in a magnetic field of 5 T obtained at different positions around the interface of the junction pictured in Fig. 1a (see Supplementary Fig. S3 for more experimental data). A strikingly unconventional interface Landau quantization is revealed in the rippled region near the interface (within several nanometers). The spectra recorded at positions 4, 5, and 6 exhibit Landau quantization of massless Dirac fermions for electrons (Fig. 2b), i.e., below the charge neutrality point $N_c$ of the junction, however, they show Landau quantization of massive Dirac fermions for holes (Fig. 2c), i.e., above the $N_c$. Moreover, the Fermi velocity for the electrons and the effective mass for the holes at the positions 4, 5, and 6 are almost identical to that obtained in Fig. 1d and Fig. 1e, respectively. It indicates that the Landau quantization at these positions switch between the characteristic of graphene monolayer and Bernal bilayer when the type of the charge carriers is changed.

We find that the spectroscopic features of the unconventional interface Landau quantization are closely linked to the quantum conductance across the interface of the monolayer-bilayer planar junction[1,6,7]. In the unipolar regime of a graphene junction consisting of the monolayer and bilayer regions, the Hall conductance plateaus across the interface are solely determined by those edge modes that permeate the entire system and follow[1,6]

$$g = \min(|v_1|, |v_2|). \quad (3)$$

Here, $v_1 = \pm 2, \pm 6, \pm 10\ldots\ldots$ is the filling factor in the graphene monolayer[13-15] and $v_2 = 0, \pm 4, \pm 8, \pm 12\ldots\ldots$ is the filling factor in the gapped bilayer[28]. Figure 3 shows the expected Hall conductivity as a function of the Fermi energy in the monolayer and bilayer regions. The Hall plateaus of the two regions are derived from their Landau quantizations in the magnetic field of 5 T, as shown in Fig. 1 and Fig. 2 (see Supplementary Fig. S4 and Fig. S5 for more analysis for the data measured in other magnetic fields). According to Eq. (3), the Hall conductance and the corresponding sequences of LLs in the junction as a function of Fermi energy are also plotted in Fig. 3. Obviously, the quantized conductance and sequences of LLs of such a junction are identical to that of graphene monolayer for electrons and becomes the same as that of Bernal bilayer for holes. In our monolayer-bilayer junction, the graphene bilayer region is directly pinned to the Rh substrate. Therefore, the tunnelling conductance in the bilayer only reflects the Landau quantization in this region itself, as demonstrated in Fig. 1 and Fig. 2. Whereas, the monolayer region (the rippled region) of the junction is not in direct contact to the Rh substrate and the tunnelling current from the STM tip to the monolayer region, especially around the interface, has to pass through the interface of the junction to the Rh substrate for the requirement of forming an electric circuit. In this case, the tunnelling spectra reflect the Hall conductivity of the junction and, consequently, we observe the unconventional Landau quantization in the rippled region around the interface (Fig. 2). For the rippled graphene about 10 nm away from the interface, there may be some contacted points between the graphene and the Rh substrate. The tunnelling current from the STM tip to this region can be partially transferred into the substrate through these contacted points, then the measured tunnelling spectra reflect the Landau quantization in the graphene monolayer (Fig. 1).

Our findings demonstrate that graphene hybrid structures may display unconventional behavior in the QH regime and that the sub-nanometre-resolved STM and the atomically well-defined interface of the planar junction allow us unprecedented opportunity to study the electronic properties of the interface. This opens the way to spatially explore the QH effect of other different graphene hybrid structures only using a STM.

**Acknowledgements**

This work was supported by the National Basic Research Program of China (Grants Nos. 2014CB920903, 2013CBA01603, 2013CB921701), the National Natural Science Foundation of China (Grant Nos. 11422430, 11374035, 11474022, 51172029, 91121012), the program for New Century Excellent Talents in University of the Ministry of Education of China (Grant No. NCET-13-0054), Beijing Higher Education Young Elite Teacher Project (Grant No. YETP0238).

**Author contributions**

L.H. conceived and provided advice on the experiment, analysis, and theoretical calculation. Y.W. performed the STM experiments. L.S.Y. analyzed the data and performed the theoretical calculations. L.H. wrote the paper. All authors participated in the data discussion.

**Additional information**

Supplementary Information accompanies this paper at http://www.nature.com/nature.

**Competing financial interests:** The authors declare no competing financial interests.

**Figure Legends**

**Figure 1 | STM image and STS of a graphene monolayer-bilayer planar junction.**
**a,** A typical atomic-resolution STM image of a graphene monolayer-bilayer planar junction on a Rh foil ($V_{sample}$ = -880 mV and $I$ = 0.35 nA). The inset shows an enlarged image in the blue frame. The hexagonal structures of graphene are overlaid onto the STM image. In the rippled region (monolayer region), we can observe honeycomb lattice, whereas, in the relatively flat area (bilayer region), only triangular lattice can be observed. **b,** Tunnelling spectra, i.e., *dI/dV-V* curves, taken at red solid circle marked in panel **a**. They show LLs of massless Dirac fermions in graphene monolayer. For clarity, the curves are offset in Y-axis and LL indices are marked. **c,** STS spectra recorded at the blue circle marked in panel **a**. The spectra show LLs of massive Dirac fermions and the LL indices are marked. The $LL_{(0,+)(1,+)}$ and $LL_{(0,-)(1,-)}$ projected on the top and the bottom graphene layer, respectively, and an energy gap between $LL_{(0,+)(1,+)}$ and $LL_{(0,-)(1,-)}$ is measured to be about 80 meV. **d**, LL peak energies recorded in the rippled region show square-root dependence on both the LL index and magnetic field, i.e., $\text{sgn}(n)(|n|B)^{1/2}$, as expected for massless Dirac fermions in graphene monolayer. The red solid line is the linear fit of the data with the slope yielding a Fermi velocity of $v_F = (1.208 \pm 0.016) \times 10^6$ m/s for electrons and $v_F = (0.992 \pm 0.014) \times 10^6$ m/s for holes. Inset: the schematic LL structure of graphene monolayer in the quantum Hall regime. **e**, LL peak energies of Bernal bilayer plotted against $+(n(n-1))^{1/2}B$ for conduction band and $-(n(n-1))^{1/2}B$ for valence band. The purple lines are the linear fit of the data with equation (2), yielding the effective mass of $(0.013 \pm 0.001)m_e$ for both electrons and holes. The inset is the schematic LLs of Bernal bilayer in the quantum Hall regime with a finite bandgap. The charge neutrality points $N_c$ of both the monolayer and bilayer regions are determined as $(73 \pm 5)$ meV and $(120 \pm 5)$ meV, respectively.

**Figure 2 | Unconventional Landau quantization around the interface of the junction. a,** STS spectra obtained at different positions (as marked in Fig. 1a) around the interface in a magnetic field of $B = 5$ T. LL indices of massless and massive Dirac fermions are marked with red and blue numbers, respectively. **b**, LL peak energies of electrons (LL energies below the $N_c$ of the junction) recorded at positions 4, 5 and 6 show linear dependence on $sgn(n)(|n|B)^{1/2}$, which corresponds to the Landau quantization of massless Dirac fermions in the graphene monolayer. The Fermi velocity of the electrons is determined as $v_F = (1.22 \pm 0.01) \times 10^6$ m/s. **c**, LL peak energies of holes (LL energies above the $N_c$ of the junction) recorded at positions 4, 5 and 6 plotted against $+B(n(n-1))^{1/2}$, which corresponds to the Landau quantization of massive Dirac fermions in the graphene bilayer. The effective mass of holes is

measured as $(0.013 \pm 0.001)m_e$. The insets of **b** and **c** show schematics of low-energy dispersion with quantized LLs in graphene monolayer and bilayer, respectively. Only the LLs plotted in solid curves are observed in the STS spectra at positions 4, 5 and 6.

**Figure 3 | The Hall conductivity in the monolayer-bilayer planar junction.** We show the expected Hall conductivity as a function of Fermi energy for the graphene monolayer and the Bernal bilayer with a finite bandgap. The Hall plateaus of the two regions are derived from their Landau quantizations in the magnetic field of 5 T (Fig. 1 and Fig. 2). The Hall conductivity of the monolayer-bilayer planar junction is also plotted according to Eq. (3). The corresponding sequences of LLs are shown in red and blue for massless Dirac fermions and massive Dirac fermions, respectively. Obviously, only LLs of massless (massive) Dirac fermions can be observed for electrons (holes).

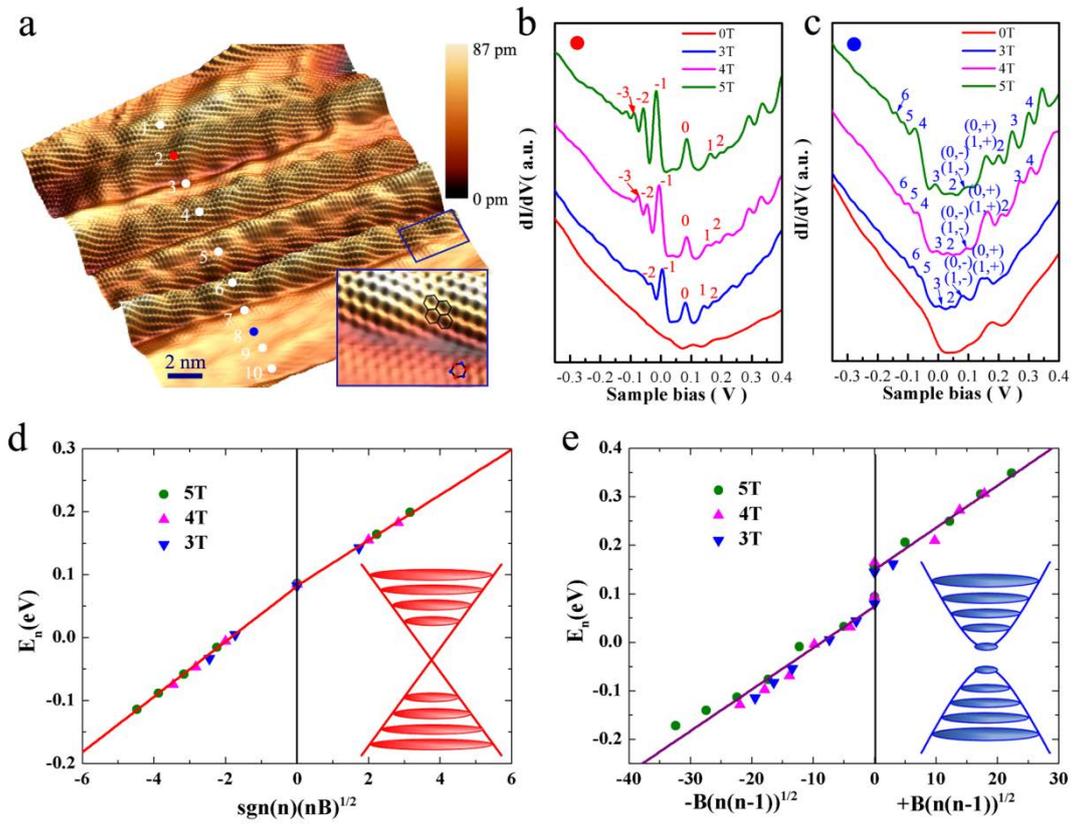

Figure 1

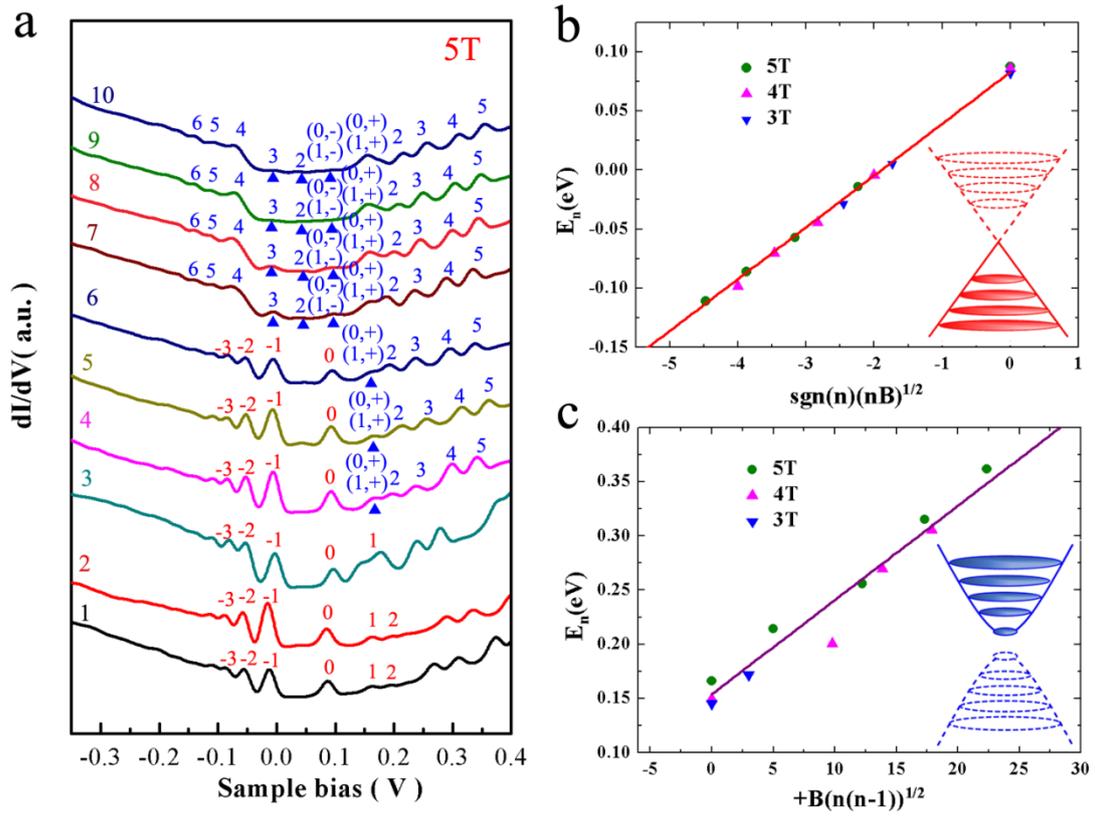

Figure 2

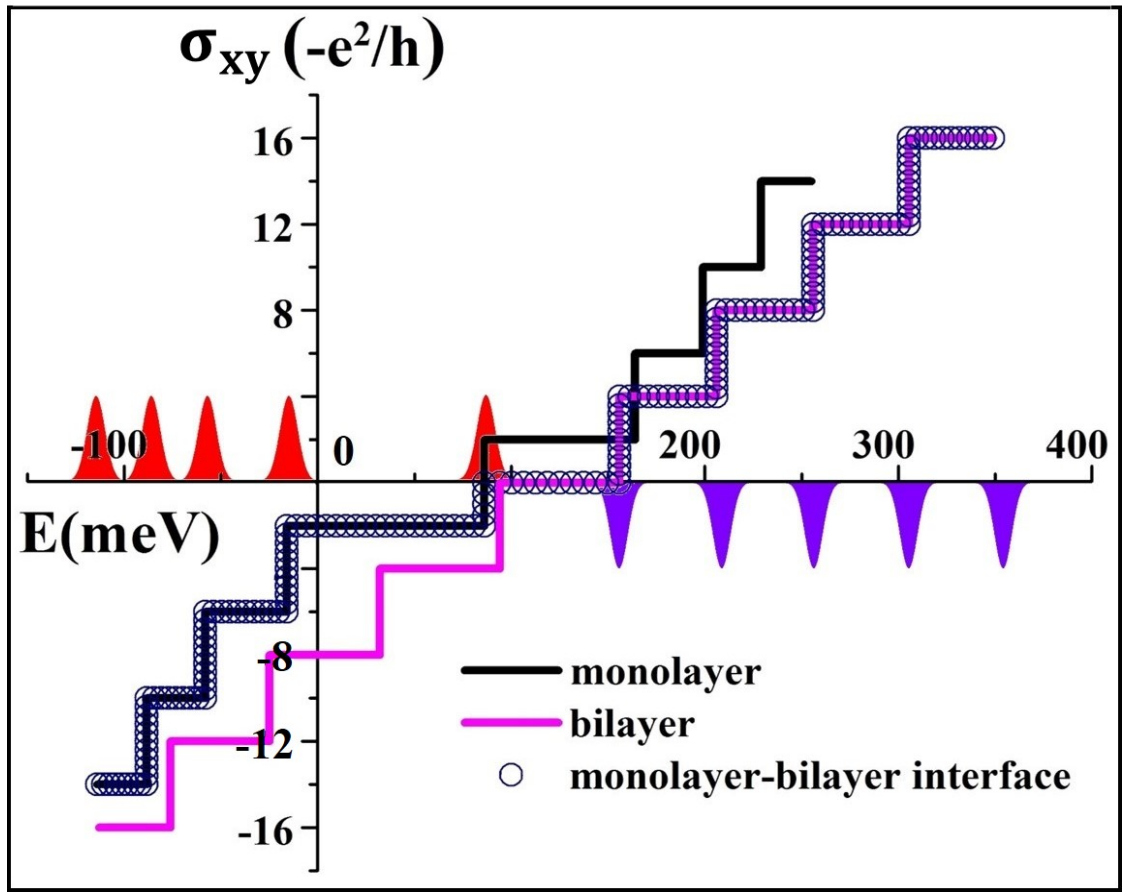

Figure 3